\documentclass[sigconf,nonacm]{acmart}

\usepackage[ruled,vlined,linesnumbered]{algorithm2e}
\usepackage{multirow}
\usepackage{makecell}
\usepackage{color, colortbl}

\AtBeginDocument{%
  \providecommand\BibTeX{{%
    \normalfont B\kern-0.5em{\scshape i\kern-0.25em b}\kern-0.8em\TeX}}}

\begin{document}

\title[{}]{A Meta-Method for Portfolio Management Using\\ Machine Learning for Adaptive Strategy Selection}

\author{Damian Kisiel}
\affiliation{%
  \institution{University College London}
  \institution{Department of Computer Science}
  \streetaddress{Gower Street}
  \city{London}
  \country{United Kingdom}}
\email{d.kisiel@cs.ucl.ac.uk}

\author{Denise Gorse}
\affiliation{%
  \institution{University College London}
  \institution{Department of Computer Science}
  \streetaddress{Gower Street}
  \city{London}
  \country{United Kingdom}}
\email{d.gorse@cs.ucl.ac.uk}

\renewcommand{\shortauthors}{}

\begin{abstract}
    This work proposes a novel portfolio management technique, the \emph{Meta Portfolio Method} (MPM), inspired by the successes of meta approaches in the field of bioinformatics and elsewhere. The MPM uses XGBoost to learn how to switch between two risk-based portfolio allocation strategies, the Hierarchical Risk Parity (HRP) and more classical Naïve Risk Parity (NRP). It is demonstrated that the MPM is able to successfully take advantage of the best characteristics of each strategy (the NRP’s fast growth during market uptrends, and the HRP’s protection against drawdowns during market turmoil). As a result, the MPM is shown to possess an excellent out-of-sample risk-reward profile, as measured by the Sharpe ratio, and in addition offers a high degree of interpretability of its asset allocation decisions.
\end{abstract}




\maketitle

\section{Introduction}
Portfolio management as we know it today has been heavily influenced by the work of Harry Markowitz \cite{Markowitz1952}, who formalized the idea of the risk-return trade-off, where investors are rewarded for taking on more risk. These portfolios, however, suffer from significant flaws \cite{Maillard2010, Ulf2006}. Institutional investors who used the standard Markowitz allocations suffered large drawdowns during recent crises; as a result, the model was criticized \cite{Roncalli2013} and risk-based portfolio allocation techniques became more popular. However, these newer methodologies, despite having rigorous underpinning mathematical structures, are not supported by any financial theory that would promote them as the optimal portfolio choice.

Therefore, rather than trying to discover a risk-based methodology that would work ideally well for all scenarios \cite{Chaves2011}, the work of this paper aims to choose the best performing strategy for given market conditions. It proposes a new portfolio construction technique called the \emph{Meta Portfolio Method} (MPM), inspired by the successes of meta approaches in the field of bioinformatics \cite{Charoenkwan2020, Ishida2008}, among other areas. The MPM combines two risk-based portfolio allocation strategies, namely the \emph{Hierarchical Risk Parity} (HRP) and more classical \emph{Naïve Risk Parity} (NRP), by allocating capital to the one predicted to provide better risk-adjusted performance in the next investment period.

\section{Background}

\subsection{Meta Methods}

The idea behind meta approaches is that the performance of a task can often be greatly enhanced by combining multiple methodologies, as opposed to using any of the constituent techniques alone. Meta approaches have been successfully applied in the field of bioinformatics \cite{Charoenkwan2020, Ishida2008}, but have not yet been widely explored in portfolio management. An exception is provided by \cite{Jaeger2021}, in which the authors did identify key features of the variance-covariance matrix that help in studying the behavior of different risk-based allocation strategies. However, this was an ex-post analysis where the model could only make retrospective decisions as to which strategy performed better. What is lacking from the existing literature is an ex-ante analysis that would be able to recommend which of the strategies should be followed in the future, given a set of statistical characteristics of a given asset universe. The research described here aims to bridge this gap.

\subsection{Risk-Based Portfolio Construction Strategies}

As mentioned in the introduction, the meta-strategy of this work combines two risk-based portfolio construction strategies, Naïve Risk Parity (NRP) and Hierarchical Risk Parity (HRP). This section will give a brief introduction to each of these, followed by a comparison of their relative strengths and weaknesses.

\subsubsection{Naïve Risk Parity (NRP)}

Risk parity portfolios aim to maximize the portfolio diversification level in the hope that a more diversified portfolio will be less vulnerable to unfavorable market conditions. The Naïve Risk Parity \cite{Roncalli2013} method is considered to be the simplest implementation of a risk-based methodology, and aims to equalize the risk contributed to the overall portfolio by each asset. Its portfolio weights can be computed as follows

\begin{equation}
w_{i}=\frac{\frac{1}{\sigma{}_{i}{}^{2}}}{\sum_{j=1}^{n}\frac{1}{\sigma_{j}{}^{2}}},
\end{equation}
where $\sigma_i^2$ denotes the variance of returns of asset $i$.

\subsubsection{Hierarchical Risk Parity (HRP)}

Hierarchical Risk Parity \cite{DePrado2016}, similar to other risk-budgeting approaches, considers the variance-covariance matrix during its construction process. However, it in addition makes use of hierarchical clustering to group together similar assets, to avoid computing correlation estimates between all pairs of assets. This process uses three steps: tree clustering, quasi-diagonalization, and recursive bisection.

In the first step the algorithm performs tree clustering by first transforming the correlation matrix into a distance matrix $D$, where each element is given by

\begin{equation}
d_{i,j}=\sqrt{\frac{1}{2}(1-\rho_{i,j})}.
\end{equation}
The newly created matrix $D$ is then transformed by computing the Euclidean distance between all column pairs as follows
\begin{equation}
\tilde{d}_{i,j}=\sqrt{\sum_{n=1}^{N}(d_{n,i}-d_{n,j})^{2}},
\end{equation}
creating a new matrix representation which gives the correlation of each asset with the rest of the portfolio. This allows the formation of a tree structure according to the $\tilde{d}_{i,j}$ ‘scores’. In the second step, quasi-diagonalization, the tree structure created in the previous step is used to reorganize the rows and columns of the correlation matrix in such a way as to make similar assets sit next to each other and investments considered to be different be placed further apart. The last step performs recursive bisection wherein the standard inverse-variance methodology is applied to the clusters defined in the previous steps. Because asset weights are assigned in a top-down fashion, according to each sub-cluster’s variance, only assets within the same cluster compete for allocation, achieving diversification benefits across all levels of the tree structure.

\subsubsection{Comparison of the Component Strategies}

\begin{figure}[h]
    \centering
    \includegraphics[width=\linewidth]{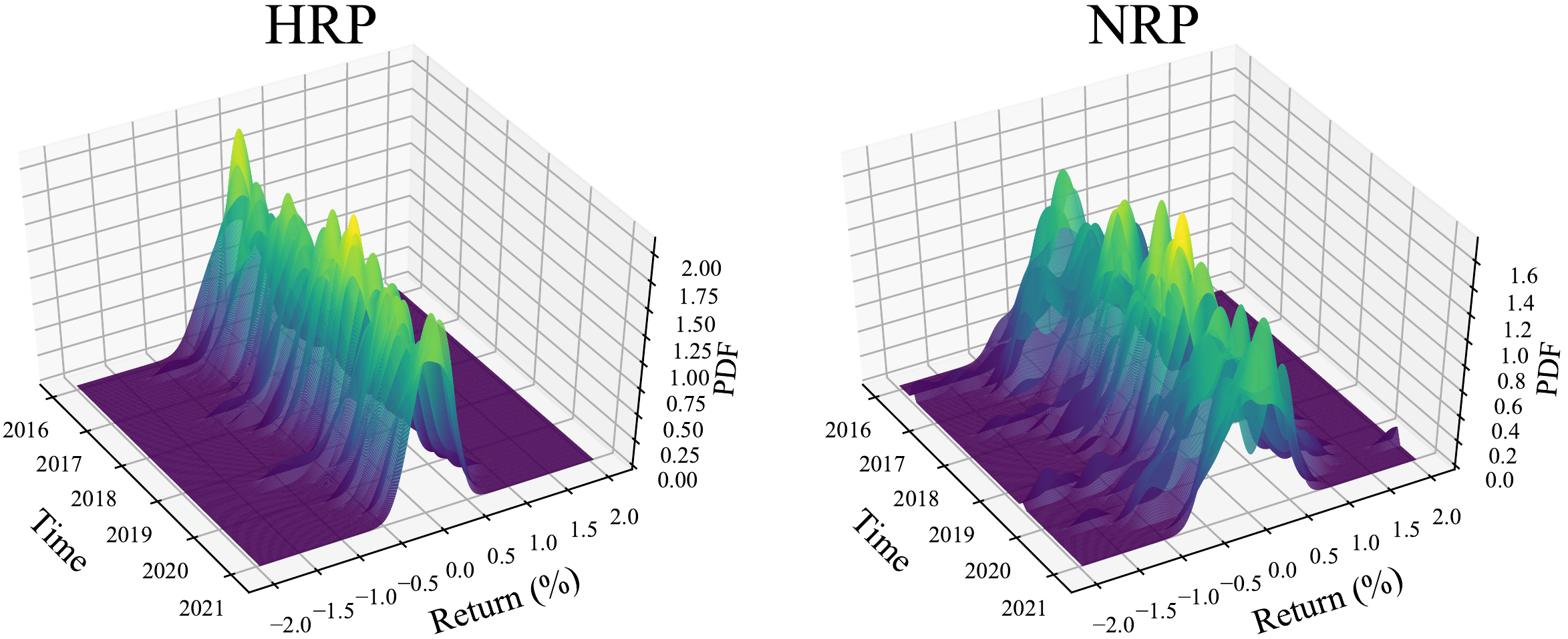}
    \caption{3D plots showing changes in the probability density function over time.} \label{fig1}
    \Description{The figure contains two 3D plots: one for the HRP strategy and the other for the NRP strategy. The plots show differences in the realised volatility profiles between the two strategies.}
\end{figure}

Each component strategy displays a different realized volatility profile, as seen in Figure~\ref{fig1}. HRP gives a portfolio with a much lower variance. It offers a return stream with heavily damped tails which makes it a safer option. Some negative tails can be spotted but these are much less common and less pronounced than those in NRP’s profile, which displays heavy tails in its return distributions. However, since fat tails also mean a higher probability of a large positive return, NRP has the potential of being considerably more profitable than HRP during certain market conditions.

\section{Methodology}

\subsection{Data Selection \& Preparation}

Results presented in this work are based on ten different investment universes, each containing five distinct asset classes. A set of 18 exchange traded funds (ETFs), as shown in Table~\ref{tab1}, is used as a basket from which assets can be drawn with replacement to construct a specific asset universe. The selection routine is constrained so as to allow each universe to share at most two ETFs with any other asset universe. The reason for selecting market indices, rather than individual assets, is that they are generally uncorrelated and hence offer higher diversification benefits. The list of ETFs shown in Table~\ref{tab1} has been chosen in such a way as to represent as many distinct asset classes as possible. Data collected on each ETF includes daily returns that span the period from November 2004 until May 2021. All assets are denominated in US Dollars, the domestic currency of the investors considered in this study.

\begin{table}
  \caption{Basket of ETFs.}
  \label{tab1}
  \begin{tabular}{c|l|c|l}
    \toprule
    \textbf{Ticker} & \textbf{Name} & \textbf{Ticker} & \textbf{Name}\\
    \midrule
    AGG & Aggregate Bonds & VNQ & Real Estate\\
    EEM & Emerging Markets & XLV & Health Care Sector\\
    EFA & Developed Markets & IWD & Value Equities\\
    EWJ & Japanese Equities & IWM & Small-Cap Equities\\
    GLD & Gold & LQD & Corporate Bonds\\
    QQQ & Nasdaq-100 Index & TLT & 20+ Year T-Bonds\\
    SHY & 1-3 Year T-Bonds & VO & Mid-Cap Equities\\
    SPY & S\&P 500 Index & VBR & Small-Cap Value\\
    TIP & I-L Bonds & XLY & Consumer Services\\
  \bottomrule
\end{tabular}
\end{table}

The original data contains daily price information; since we are interested in the growth of each universe component, the simple (arithmetic) return from time $t-1$ to time $t$ is computed as follows

\begin{equation}
r_{n,t}=\frac{P_{n,t}}{P_{n,t-1}}-1,
\end{equation}

\noindent where $P_{n,t}$ is the price of asset $n$ at time (day) $t$. If we have $N$ assets in a portfolio and their weights at time $t-1$ are given by $w_{1,t-1}, w_{2,t-1}, \ldots, w_{N,t-1}$, the portfolio returns are computed using

\begin{equation}
R_{t}=\sum_{n=1}^{N}w_{n,t-1}r_{n,t}.
\end{equation}

\noindent In order to simplify calculations, it is assumed the risk-free return is zero throughout the investment period and hence the return above is our variable of interest.

\subsubsection{Computation of Return Covariance Matrix}

Both of the component strategies (HRP and NRP) require an estimate of the variance-covariance matrix between each pair of asset return series. This matrix will be used both for the computation of the features to be input to the machine learning model (next section) that selects the strategy to be used in the next time period, and for the internal construction of portfolio weights. The multivariate GARCH Dynamic Conditional Correlation (DCC) model \cite{Engle2002} is employed here to compute the variance-covariance matrix. The DCC model is well-regarded because it allows for time-varying correlations between assets.

\subsection{Computation of Features}

The set of predictive features includes measures that aim to describe each constituent strategy’s most recent performance, such as average return, realized volatility \cite{Izzeldin2019}, maximum drawdown \cite{Chekhlov2005} and downside deviation \cite{Sortino1994}, as well as a given investment universe’s mean return and standard deviation.

In addition, more elaborate transformations of the variance-covariance matrix are carried out. For example, \emph{meanCORR} and \emph{stdCORR} compute the average and the standard deviation, respectively, of all lower triangular elements of the correlation matrix and indicate overall level and heterogeneity of correlations at a given point in time. Other features include the \emph{non-parametric k-nearest neighbor entropy estimator} \cite{Lombardi2016}, the \emph{quality ratio} \cite{Butler2018}, that aims to capture how much diversification benefit each investment universe is able to offer, and the \emph{standardized generalized variance} \cite{Najarzadeh2019}, which measures the overall volatility of a specific asset universe.

As described in section 2.2, the two component strategies (HRP and NRP) differ in the way they are constructed. Since the HRP considers an additional clustering step, its performance depends on the level of the hierarchical structure present in a given asset universe, measured by the \emph{cophenetic correlation coefficient} \cite{Sokal1962}, with the quality of clustering computed using the intra-cluster variance \cite{Halkidi2001}. Finally, a set of features characterizing some properties of the correlation matrix are calculated, including its determinant, condition number \cite{Cline1979} and the fraction of variance that can be explained by the eigenvalues that lie outside the Marchenko-Pastur distribution \cite{Marchenko1967}.

\subsection{Outline of the Meta Portfolio Method (MPM)}

\begin{figure}[h]
    \centering
    \includegraphics[width=\linewidth]{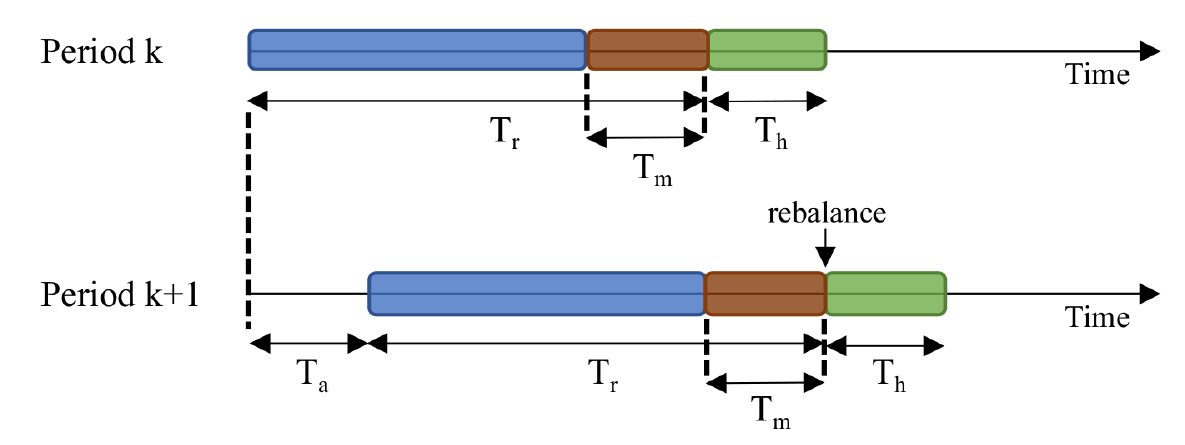}
    \caption{Meta Portfolio Method (MPM) timeline.} \label{fig2}
    \Description{The figure represents the two-period sequential outline of the Meta Portfolio Method (MPM). In the second period all moving windows are advanced by length Ta.}
\end{figure}

The sample period denoted by $T_r$ in Figure~\ref{fig2} is used as input to the multivariate GARCH DCC model that computes the variance-covariance matrix, which is then used in the construction of the two component strategies and to compute a set of model features. Observations collected during the brown sample period (denoted by $T_m$) in Figure~\ref{fig2} are used to calculate features that describe each component strategy’s most recent performance. Once all model features have been computed, the strategy selection process can take place.

At the beginning of the green band (denoted by $T_h$) in Figure~\ref{fig2} the machine learning model of the next section decides which of the two component strategies (HRP or NRP) should be followed in the next investment period. All available capital is then invested in the selected strategy and held during the holding period of length $T_h$ and its returns recorded. In the next investment period (Period $k+1$) the rolling windows are advanced by the time interval $T_a$ and the two component strategies are separately rebalanced. Then, the process is repeated. All returns of the Meta Portfolio Method are computed in a walk-forward manner so that the strategy selection algorithm has access only to the historical data, and transaction costs are taken into account when evaluating performance. The whole procedure is summarized in Algorithm 1, below.

\noindent

\begin{algorithm}
\SetAlgoLined
import daily asset returns\\
initialise $T_m, T_r, T_h$ and $T_a$\\
\While{t $<$ last observation}{
	compute var-cov matrix on $T_r$ data using GARCH DCC\\
	construct the two component strategies\\
	compute XGBoost features\\
	\For{$i\gets1$ \KwTo $T_m$}{
    		compute performance-related features
    	}
	apply Bayesian hyperparameter optimization\\
	predict the Sharpe ratio spread using optimized model\\
	\eIf{$spread>0$}{
   		select HRP strategy\;
   	}{
   		select NRP strategy\;
  	}
  	\For{$i\gets1$ \KwTo $T_h$}{
    		compute returns generated by the MPM
    	}
    	advance the sliding window by $T_a$
}
\caption{Meta Portfolio Method (MPM)}
\end{algorithm}

\subsection{Training Methodology}

\subsubsection{Target Variable}

Since we are interested in predicting which of the two strategies is going to perform better (on a risk-adjusted basis) during the next investment period, the target variable is the \emph{Sharpe ratio spread}, the difference between Sharpe ratios of the HRP and NRP strategies, given by

\begin{equation}
y_{target}=SR_{HRP}-SR_{NRP}.
\end{equation}

The process of switching between the two component strategies can be formulated as a binary classification problem, since at any point in time we only have two choices at our disposal. However, we notice from equation 6 that the target variable is a real-valued number, which allows us to work in a regression setting. The advantage of this is that the magnitude of the Sharpe ratio spread carries additional information that a machine learning model can leverage during the training process. All real-valued predictions can then be converted to a binary allocation decision using the HRP strategy if $y_{predicted}\geq0$, and the NRP strategy otherwise.

\subsubsection{ML Model and Hyperparameter Optimization}

XGBoost \cite{Chen2016} is selected to solve the regression problem above. The main advantage of XGBoost, in this problem setting, comes from the fact that it is able to construct non-linear relations between the input features while, at the same time, preserving a high degree of interpretability through its built-in feature importance functions, an important consideration when making portfolio allocation decisions. However, to fully harness the predictive power of XGBoost it is necessary to carry out careful hyperparameter optimization. Each set of model-building data ($T_r$ sections in Figure~\ref{fig2}) is split into a training part which represents 70\% of observations and a test part which contains the remaining 30\% of data. Only the training data is used for hyperparameter tuning purposes, with 5-fold cross-validation implemented to get the average regression scores across all folds. Due to the large number of hyperparameters, a classical grid search would be too time-consuming, and Bayesian hyperparameter optimization is therefore used instead to select the best set of hyperparameters for each asset universe.

\section{Results}

\subsection{Cumulative and Risk-Adjusted Returns}

Table~\ref{tab2} gives a performance comparison between the Meta Portfolio Method (MPM) and its constituent strategies (HRP \& NRP). The MPM shows the strongest performance in all investment universes, with Sharpe ratio improvements over the HRP and NRP having p-values of $6.01\times{10}^{-9}$ and $2.45\times{10}^{-7}$, respectively, and cumulative return improvements having comparable p-values of $1.53\times{10}^{-7}$ and $4.50\times{10}^{-4}$. The improvements provided by the MPM are therefore clearly statistically significant.

\begin{table}
\centering
\caption{Performance comparison.}\label{tab2}
\begin{tabular}{>{\centering}m{1.65cm}|>{\centering}m{1.2cm}|>{\centering}m{1.2cm}|>{\centering}m{1.2cm}|>{\centering}m{1.2cm}}
\hline
 & \makecell{\textbf{Asset} \\ \textbf{Universe}} & \textbf{HRP} & \textbf{NRP} & \textbf{MPM}\tabularnewline
 
\hline
\multirow{10}{*}{Sharpe Ratio} & 1 & 1.37 & 1.25 & 1.64\tabularnewline
\cline{2-5} \cline{3-5} \cline{4-5} \cline{5-5} 
 & 2 & 1.20 & 1.17 & 1.32\tabularnewline
\cline{2-5} \cline{3-5} \cline{4-5} \cline{5-5} 
 & 3 & 1.35 & 1.23 & 1.82\tabularnewline
\cline{2-5} \cline{3-5} \cline{4-5} \cline{5-5} 
 & 4 & 1.07 & 1.03 & 1.22\tabularnewline
\cline{2-5} \cline{3-5} \cline{4-5} \cline{5-5} 
 & 5 & 1.30 & 1.38 & 1.59\tabularnewline
\cline{2-5} \cline{3-5} \cline{4-5} \cline{5-5} 
 & 6 & 1.68 & 1.34 & 1.87\tabularnewline
\cline{2-5} \cline{3-5} \cline{4-5} \cline{5-5} 
 & 7 & 1.67 & 1.44 & 2.31\tabularnewline
\cline{2-5} \cline{3-5} \cline{4-5} \cline{5-5} 
 & 8 & 1.12 & 1.07 & 1.41\tabularnewline
\cline{2-5} \cline{3-5} \cline{4-5} \cline{5-5} 
 & 9 & 1.01 & 0.99 & 1.23\tabularnewline
\cline{2-5} \cline{3-5} \cline{4-5} \cline{5-5} 
 & 10 & 0.84 & 0.87 & 1.05\tabularnewline
\hline
\multicolumn{5}{c}{}\tabularnewline
\hline 
\multirow{10}{*}{\makecell{Cumulative \\ Return (\%)}} & 1 & 9.06 & 15.79 & 16.37\tabularnewline
\cline{2-5} \cline{3-5} \cline{4-5} \cline{5-5} 
 & 2 & 60.50 & 74.84 & 82.98\tabularnewline
\cline{2-5} \cline{3-5} \cline{4-5} \cline{5-5} 
 & 3 & 9.31 & 16.47 & 17.48\tabularnewline
\cline{2-5} \cline{3-5} \cline{4-5} \cline{5-5} 
 & 4 & 66.23 & 95.52 & 116.60\tabularnewline
\cline{2-5} \cline{3-5} \cline{4-5} \cline{5-5} 
 & 5 & 42.39 & 68.46 & 73.18\tabularnewline
\cline{2-5} \cline{3-5} \cline{4-5} \cline{5-5} 
 & 6 & 11.19 & 18.05 & 19.73\tabularnewline
\cline{2-5} \cline{3-5} \cline{4-5} \cline{5-5} 
 & 7 & 9.18 & 20.92 & 23.84\tabularnewline
\cline{2-5} \cline{3-5} \cline{4-5} \cline{5-5} 
 & 8 & 64.89 & 84.11 & 107.14\tabularnewline
\cline{2-5} \cline{3-5} \cline{4-5} \cline{5-5} 
 & 9 & 54.05 & 79.85 & 97.23\tabularnewline
\cline{2-5} \cline{3-5} \cline{4-5} \cline{5-5} 
 & 10 & 22.50 & 51.00 & 60.12\tabularnewline
\hline
\end{tabular}
\end{table}

\begin{figure}[h]
    \centering
    \includegraphics[width=\linewidth]{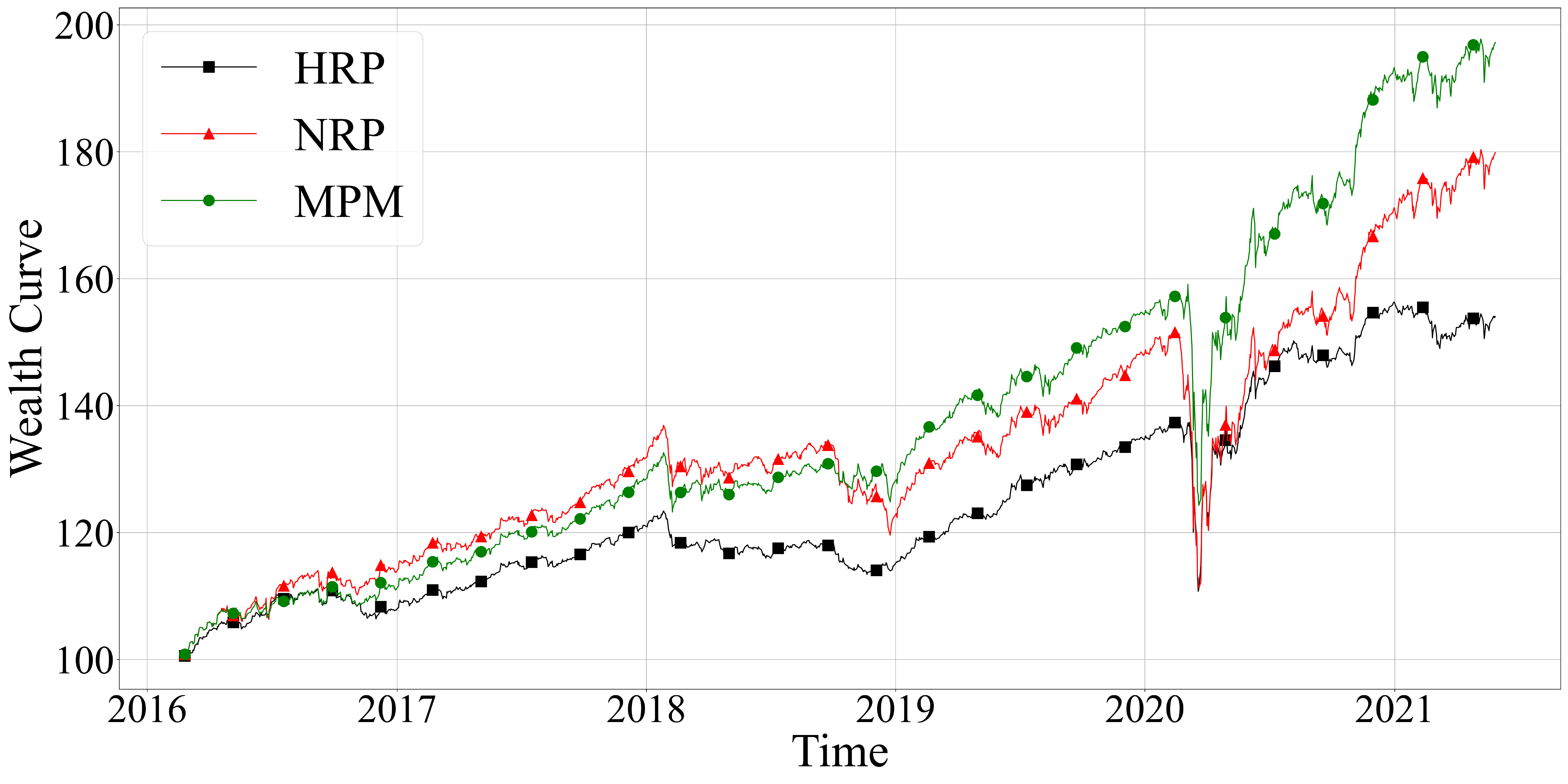}
    \caption{Comparison of cumulative returns for investment universe 9.} \label{fig3}
    \Description{Plot showing performance comparison between the MPM, HRP and NRP strategies for one example asset universe. The MPM achieves the highest cumulative return, followed by NRP and then HRP.}
\end{figure}

Figure~\ref{fig3} shows an example wealth curve for investment universe 9 (a middle-performing example), over the entire investment horizon. The MPM (green circle) is able to harness the fast growth opportunity offered by the NRP (red triangle) during market uptrends while at the same time offering a good level of protection presented by the HRP (black square) during more turbulent periods, such as the onset of the COVID-19 pandemic in 2020. It is by such means the MPM achieves the highest cumulative returns and Sharpe ratios, as evidenced by the results in Table~\ref{tab2}.

\subsection{Feature Importance}

\begin{figure}[h]
    \centering
    \includegraphics[width=\linewidth]{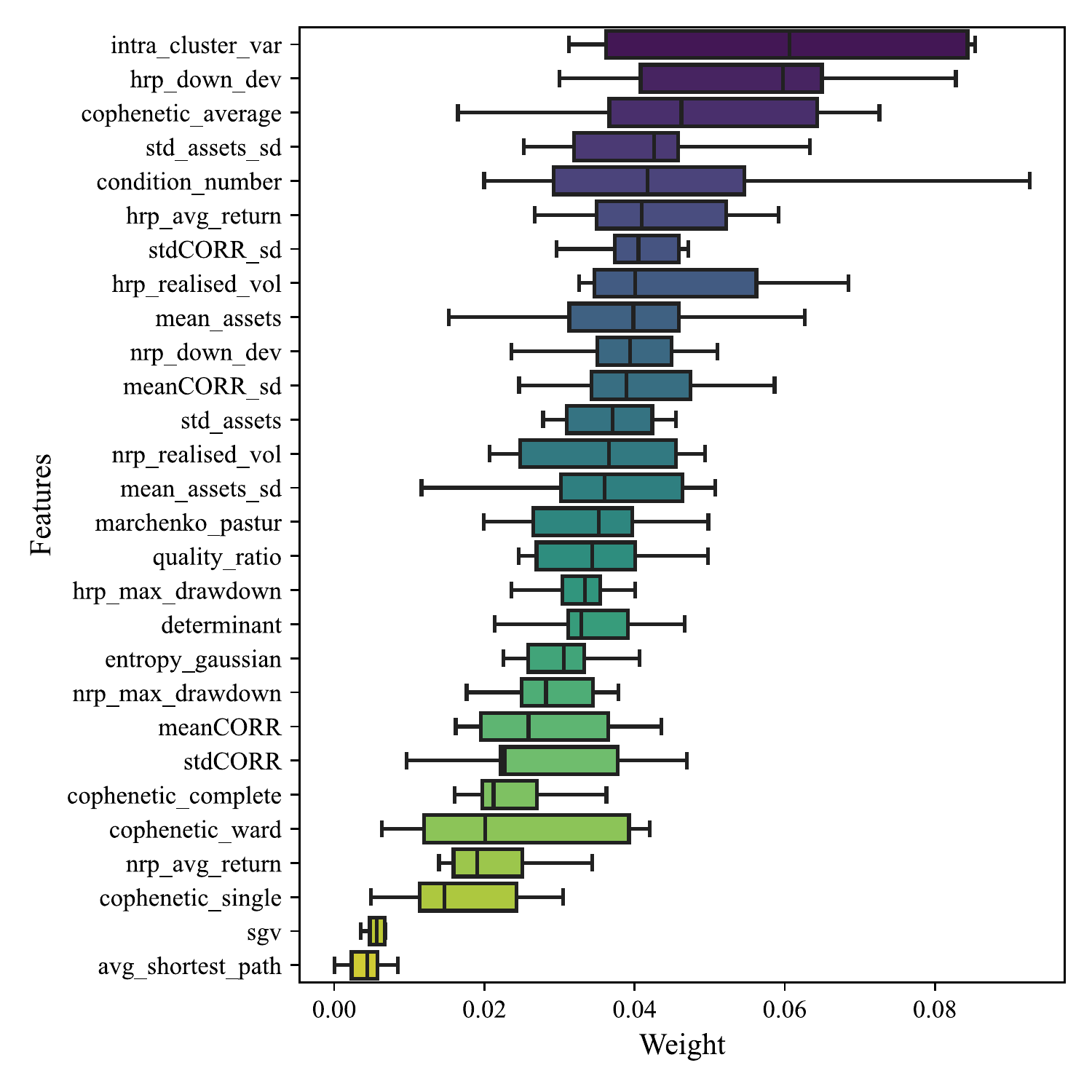}
    \caption{Boxplot with feature importances.} \label{fig4}
    \Description{Boxplot showing the relative importance of the model features averaged across all asset universes. Features are arranged from most important to least important.}
\end{figure}

One benefit of XGBoost \cite{Chen2016}, as previously mentioned, is that it is relatively easy to extract information about feature importance, which allows a better understanding of what is driving the output of the MPM. Figure~\ref{fig4} shows a boxplot of the importance scores for each feature across all investment universes. We notice that among the top three features is ‘hrp\_down\_dev’, the downside deviation of the HRP strategy. This makes intuitive sense since this feature informs the model about the historical downside risk experienced by the HRP strategy and hence the overall risk at any point in time. What is less obvious, however, is the high importance score assigned to ‘intra\_cluster\_var’ and ‘cophenetic\_average’. These features correspond to the intra-cluster variance and the cophenetic correlation coefficient, respectively, and, as described earlier, quantify the level and the complexity of the hierarchical structure in a given investment universe. However, since this information is taken into account when constructing the HRP strategy, it is understandable these two features play a big role in predicting the relative performance of the HRP strategy.

\section{Conclusions}

This work has introduced the Meta Portfolio Method (MPM), which uses XGBoost to combine two different portfolio allocation paradigms, the Hierarchical Risk Parity (HRP) and the Naïve Risk Parity (NRP), both frequently used in practice, into a single portfolio allocation framework. The MPM is able to take advantage of both the fast growth opportunities offered by the NRP in favorable market conditions, and the protection against large drawdowns during market downturns offered by the HRP. As a result, the MPM gives substantially higher cumulative returns than either of the component strategies individually, while also maintaining the highest level of risk-adjusted return, as measured by the Sharpe ratio. Moreover, the use of XGBoost within the MPM results in an easily interpretable model, desirable to industry practitioners.

The performance of the MPM strategy has been analyzed on ten different multi-class asset universes, giving confidence the MPM is not overfitted to a particular data set. It should be noted it would not be possible to construct significantly more asset universes based only on ETFs. However, while ETFs are desirable in that these assets can be chosen to be less correlated in performance than, say, individual equities, it is not an absolute necessity to use them, and future work will look at the alternative of having larger asset universes composed of less weakly correlated components.

\bibliographystyle{ACM-Reference-Format}
\bibliography{refs}

\end{document}